# STUDY OF PROXIMITY EFFECTS IN SUPERCONDUCTOR/FERROMAGNET INTERFACE USING WAVEGUIDE ENHANCEMENT OF NEUTRON STANDING WAVES

V.L. Aksenov, K.N. Zhernenkov, Yu. N. Khaidukov, Yu.V. Nikitenko
*Frank Laboratory of Neutron Physics, JINR, Russia*

L. Bottyán, F. Tanczikó, D. Merkel, B. Nagy, L. Deák, D.L. Nagy, E. Szilágyi
*KFKI Research Institute for Particle and Nuclear Physics of the HAS, Budapest, Hungary*

Zs.E. Horváth
*Research Institute for Technical Physics and Materials Science of the HAS, Budapest, Hungary*

L. Kiss
*Research Institute for Solid State Physics and Optics of the HAS, Budapest, Hungary*

A. Csík, K. Vad
*Institute of Nuclear Research of the HAS, Debrecen, Hungary*

G. Langer
*Department of Solid State Physics, University of Debrecen, Debrecen, Hungary*

A. Rühm
*Max-Planck-Institut für Metallforschung, Stuttgart, Germany*

Due to their incompatible nature, singlet superconductivity (S) and ferromagnetic (FM) order do not coexist. The exchange field, in a magnetically ordered state, tends to align spins of Cooper pairs in the same direction, thus preventing a pairing effect [1]. Conversely, ferromagnetic ordering is unlikely to appear in the superconducting phase. The energy for ferromagnetic ordering decreases and, instead of ferromagnetism, a non-uniform magnetic ordering ('cryptoferromagnetism') may appear in bulk materials [2]. However, due to the great progress of high-quality hybrid S/FM preparations, coexistence of S and FM can easily be realized in thin film heterostructures. On the one hand, Cooper pairs can penetrate the FM layer and in the interface region of a few nanometers may induce superconductivity even in the presence of a relatively large exchange field. On the other hand, magnetic order penetrates the SC layer and various novel FM and SC states may form with spatial oscillations and non-monotonic temperature variations with promising novel applications of structures like π-Josephson junctions, and S/FM spin-valves.

An S layer has been reported to affect the magnetic properties of the FM with scenarios – beside cryptoferromagnetism [2-4]– of magnetization leakage from FM into S layer [5,6], as well as a change of indirect exchange coupling of neighboring FM layers through S layer [7]. It is rather difficult to experimentally verify these theoretical predictions. This is why only a few experiments have been performed to study the influence of superconductivity on FM. In Refs. [3,4], for example, the observed reduction of magnetization below $T_c$ in an S/FM bi-layer was interpreted as a cryptoferromagnetic effect, however, the experiment could also be interpreted as a consequence of magnetization leakage (an 'inverse proximity effect'), namely, by an induced negative magnetization in S layer and a suppression of the magnetization in the FM layer.

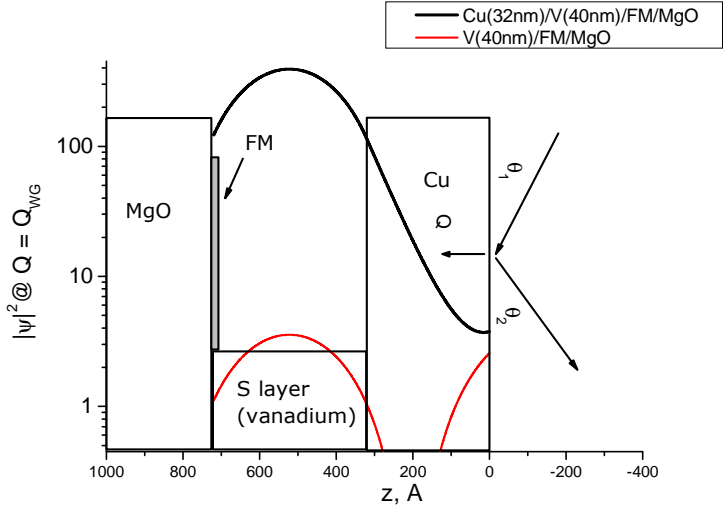

Fig. 1. Density of neutron flux inside a waveguide structure Cu(32 nm)/V(40 nm)/57Fe(0.5-1.5 nm)/MgO with and without Cu layer at Q = 0.009 Å-1 (solid black and dashed red lines, resp.). SLDs of materials are schematically shown by rectangular and incoming and reflected beams are shown by arrows.

In order to judge upon the interpretations one has to perform experiments with methods of depth resolution matching the layer thicknesses, like Polarized Neutron Reflectometry (PNR). PNR is sensitive to the magnetic depth profile $\vec{M}(\vec{r})$. In order to be able to observe such proximity effects, samples of layer systems with ultra-thin FM layer and possibly with reduced exchange coupling strength (as manifested by a decreased Curie temperature $T_{Curie}$) are necessary. Simulations reveal that a straighforward usage of conventional PNR, it is impossible to detect such weak magnetic signals. In order to increase the magnetic signal the enhanced neutron standing wave ('waveguide') regime is to be used. In order to enhance standing waves, we place the S/FM interface between two layers with high scattering potentials. Such optimization study has been performed for several candidate materials [8]. Accordingly, using the Cu(320 Å)/V(400 Å)/$^{57}$Fe(5-15 Å)/MgO structure, one can increase the neutron flux by two orders of magnitude (Fig.1) near the S/FM interface (at $Q = Q_{WG}$ (0.009 Å$^{-1}$ in the present case). Such a flux increase leads then to an enhancement of quasi-secondary radiations (spin-flip scattering and diffuse scattering) at the same value of $Q$.

Several Cu/V/$^{57}$Fe/MgO(001) samples (size ~20×10×2 mm$^3$) with different thinkness of, and different exchange field in the FM layer (the latter achieved by co-evaporation of V with $^{57}$Fe in different proportions) have been prepared in the KFKI Research Institute for Particle and Nuclear Physics, Budapest by molecular beam epitaxy (MBE) and, using magnetron-sputtering, further multilayers have been prepared in the Institute of Nuclear Research in Debrecen. To check structural, magnetic and superconducting properties of the samples, various experimental techniques were used, such as small- and wide-angle X-ray scattering, Rutherford backscattering (RBS) (of helium ions), Secondary Neutral Mass Spectrometry (SNMS) and SQUID magnetometry. X-ray Measurements have shown good quality of majority of the samples with layer structure close to nominal. For example, in Fig. 2$a$ an x-ray reflectogram of the sample Cu(32 nm)/V(40 nm)/Fe(0.5 nm)/MgO is shown. The curve shows a total reflection region and clearly resolved Kiessig oscillations giving an evidence of a well layered structure with small interface roughness. RBS and SNMS of samples with all layers grown at 300 C by MBE reveal a considerable mixing of the Cu layer with the V underlayer. Samples therefore have been re-grown at 300 C (Fe and V layers) and cooled down before Cu layer growth, eliminating the mixing.

Magnetic and superconducting properties of the samples were analysed using a Quantum Design SQUID magnetometer at the Research Institute for Solid State Physics and Optics, Budapest. All measured samples showed a finite magnetization at 5 K (see for example Fig. 2b). Saturation magnetization and coercivity of the FM layer at $T \sim 10$ K were found to be $M_{sat} = [0.2\text{-}1.4]$ kOe and $H_c = [0.04\text{-}0.18]$ kOe, respectively.

The superconducting transition temperature $T_C$ of the S layer was found by taking temperature scans of the magnetization in the range [1.8-10] K in a magnetic field of $H = 10$ Oe. A change of the magnetization at $T \approx 3$ K was observed for several samples. The absence of superconductivity for the rest can be explained by a substantial intermixing of Cu and V at the Cu/V interface or its suppression by large exchange coupling in the FM layer.

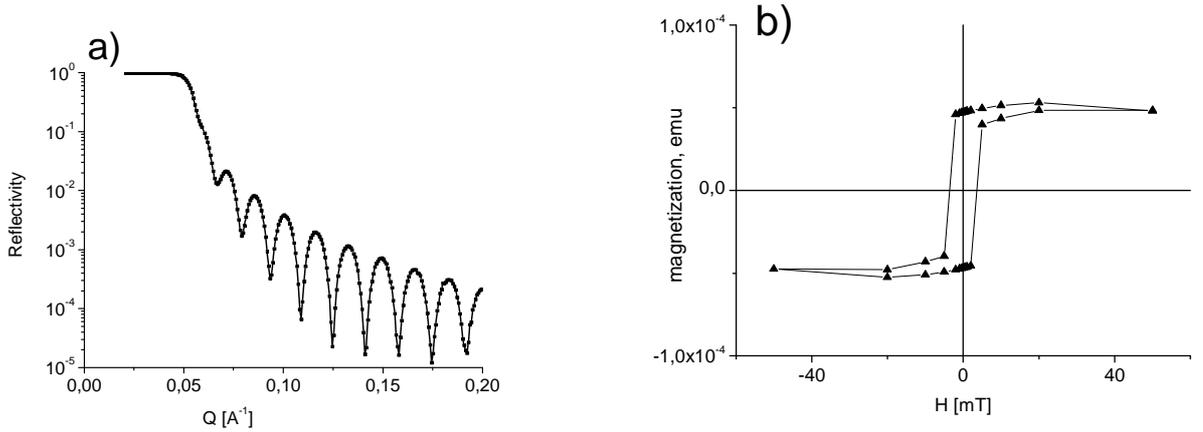

Fig. 2. Characterization curves for the sample Cu(32 nm)/V(40 nm)/Fe(1 nm)/MgO: a) X-ray reflectivity curve, using Cu K$_\alpha$; b) Hysteresis loop measured by SQUID at T = 10 K.

In order to check the waveguide properties of the samples, room-temperature measurements were performed by polarized neutrons at the NREX+ reflectometer [9] of FRM-II, Munich, Germany. The presence of waveguide regime was searched for in the spin-flip channel. Spin-flip scattering originates from magnetization in FM layer non-collinear with the direction of the external magnetic field $H$. In order to induce such a magnetization in our samples, the samples were magnetized to saturation then rotated by 90° for the reflectometric experiment in a magnetic field, $H$, with $0 < H < H_c$. In Fig. 3, polarized neutron reflectivities in the different spin states are shown in black ($R^{++}$) and red ($R^{+-}$).

The presence of the waveguide mode is proved by the dip in the $R^{++}$ and the peak in the $R^{+-}$ reflectivity curves at $Q = 0.008$ Å$^{-1}$, at a value closed to the expected one. The change in the relative intensity at the dip and the peak are 74% and 10%, respectively. By fitting the reflectivity curve using the FitSuite code [10], the absolute value of the magnetization and the angle between the external field and the magnetization of the FM layer were found to be M = 0.9 ± 0.2 kOe and 35° ± 5°, respectively. The intensity of the peak and the dip are very sensitive to the change of the magnetization vector. Our calculations show that using the present experimental setup (intensity of incoming beam, resolution) and the above described multilayer structure, it is possible to observe as low as 1% change in the FM layer magnetization. This is why such systems are especially suitable for studying the weak effects exerted by the superconducting film upon the FM layer in proximity to it below the temperature of superconducting transition.

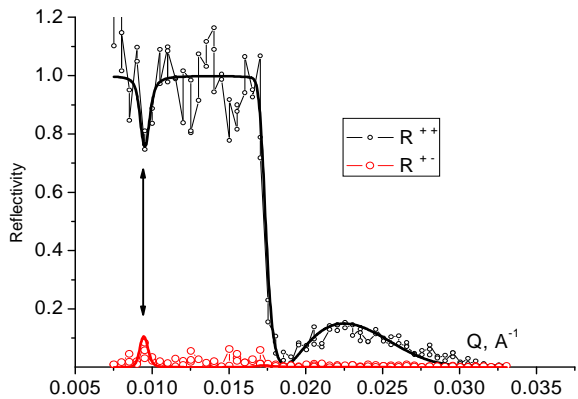

Fig. 3. Neutron reflectivities for different scattering channels from the sample Cu(32 nm)/V(40 nm)/Fe(1 nm)/MgO measured at RT in a magnetic field of 12 Oe. Arrow shows position of waveguide mode.

In conclusion, strong waveguide enhancement of neutron standing waves have been forseen by simulation and observed in various Cu/V/Fe/MO layer structures. The enhancement is large enough to detect weak magnetic signal of the thin (~ 1 to 4 nm) ferromagnetic Fe layer and changes in it by possible inverse proximity effects due to the superconducting transition of the V layer in proximity. Two series of samples were prepared by molecular beam epitaxy and by magnetron sputtering. The samples were well characterized by different techniques. Room-temperature PNR measurements proved that the well layered samples show waveguide properties, however, in case of layer mixing, the extended interface region decreases wave guide enhancement and the superconducting temperature of the sample simultaneously. The characterized samples with step-like depth profile are suitable for the scheduled PNR measurements at helium temperature.

This work was supported by the Hungarian Academy of Sciences, the Hungarian National Fund (OTKA) and the National Office for Research and Technology of Hungary under Contract Nos. MTA-DUB/01, K62272 and NAPVENEUS'05, respectively. One of the authors (Yu. Kh.) gratefully acknowledges the financial support by The Foundation for Assistance to Small Innovative Enterprises of Russian Federation (grant № 8455 UMNIK-08-3)